\documentclass[sigconf]{acmart}
\makeatletter
\renewcommand\@formatdoi[1]{\ignorespaces}
\makeatother
\setcopyright{rightsretained}
\copyrightyear{2022}
\acmYear{2022}
\acmDOI{}

\acmConference[CHI 22 Adaptive UIs Workshop]{CHI 22}{April 30th,
  2022}{New Orleans, LA}
\acmBooktitle{CHI '22 Workshop on Adaptive User Interfaces April 2022, New Orleans, LA} 
\acmISBN{}

\usepackage{natbib} 

\begin{document}

\title{A Contextual Framework for Adaptive User Interfaces}
\subtitle{Modelling the Interaction Environment}

\author{Mateusz Dubiel}
\authornote{Both authors contributed equally to this research.}
\author{Bereket Abera Yilma}
\authornotemark[1]
\author{Kayhan Latifzadeh}
\author{Luis A. Leiva}
\email{name.surname@uni.lu}
\affiliation{%
  \institution{University of Luxembourg}
  \country{Luxembourg}}

\renewcommand{\shortauthors}{M. Dubiel et al.}

\begin{abstract}
The interaction context (or \emph{environment}) is key to any HCI task 
and especially to adaptive user interfaces (AUIs),
since it represents the conditions under which users interact with computers.
Unfortunately, there are currently no formal representations to model said interaction context. 
In order to address this gap, we propose a contextual framework for AUIs
and illustrate a practical application using learning management systems as a case study. 
We also discuss limitations of our framework and offer discussion points about the realisation of truly context-aware AUIs.

\end{abstract}



\maketitle

\section{Introduction}

Research in Adaptive User Interfaces (AUIs) has provided empirical evidence about the importance of contextual information from the user's environment to aid personalisation of UIs~\cite{shankar2007user}. Although there are several works on context-aware AUIs that explore the role of said environment, this research area remains open. In practice, users evolve in a physical space together with other people and devices that often tend to influence the behaviour of the users. Additionally, a physical space by itself has certain rules that govern user behaviour and impose additional constraints on how users interact with computers, which ultimately determines how interfaces should be adapted. Nevertheless, classic approaches to AUIs focus on cues and signals solely collected from users to personalise interfaces, largely ignoring contextual information. In this work we propose a general framework which allows to incorporate the user's interaction context (or \emph{environment}) in AUIs by design.  

In order to provide contextual grounding for our framework we first briefly overview related work on AUIs (\autoref{relevant work}). Then, in \autoref{framework}, we formalise the problem by taking a systems theory perspective~\cite{Maier1996}, and present our proposed framework inspired by a classic reinforcement learning architecture~\cite{watkins1992q,sutton1998introduction}. In \autoref{case study} we present a case study that illustrates a potential application of our framework. We finish with implications of our proposal and provide several points for discussion and future work (\autoref{implications}).

\section{Related Work}
\label{relevant work}

Many studies in AUIs explore the role of user's feedback on interface personalisation. 
Some types of feedback collected include: using natural language programming to optimise behaviour of artificial agents and task disambiguation~\cite{jiang2019orc}; introducing head-gesture analysis to facilitate hands-free interaction with Head Mounted displays~\cite{yan2018headgesture};  enabling silent-speech interaction mode with a mobile device by analysing user's mouth opening degree~\cite{sun2018lip}; utilising users' mouse behaviour to automatically modify layout of a website~\cite{leiva2012automatic}; or using user body postures and physical activity to adjust the features of the surrounding environment such as ambient light, temperature or music~\cite{wang2019flextouch}.

Another line of research focused on system features that can be used for interface adaptation. 
For instance, Jiang et al.~\cite{jiang2019orc} proposed a novel layout method that adds OR-constraints (ORC) to standard constraint-based layout specifications. The proposed method unifies grid layout and flow layout, offering new possibilities for flexible UIs that were not supported by any other layout method. 
In another study, Swearngin et al.~\cite{swearngin2020scout} proposed the Scout system to help designers rapidly explore alternative interface layouts through mixed-initiative interaction with high-level constraints (e.g. semantic structure, emphasis, order) and design feedback.

\subsection{AUI Frameworks}

Previous formalizations of AUIs have approached the adaptation problem from different angles. 
One notable example in this area is the \textsc{Supple} system~\cite{gajos2004supple} which framed UI adaptation as an optimisation problem. \textsc{Supple} minimises user's effort by providing interface adaptations that meet device and user capabilities. However, \textsc{Supple} did not account for the modeling of environment features.

Similarly, Bouzit et al.~\cite{bouzit2017pda} formalised a design space for user interface adaptation, alas without modelling the environment. The proposed framework is based on Perception-Decision-Action cycle that is augmented by Learning-Prediction-Action, allowing for UI designs that are descriptive, comparative, and generative.

On the other hand, Abrah{\~a}o et al.~\cite{abrahao2021model} proposed a conceptual reference framework for intelligent user interface adaptation with a set of conceptual adaptation properties that are crucial for model-based AUIs. While the authors mention an ``Environmental Model'' as a part of their ``Context Model'' that affects system adaptation, there is no formulation provided regarding the elements that define the environment and the role of their interactions.

Overall, while previous work on proposed conceptual modelling of the system, to the best of our knowledge, conceptual modelling of the environment has not been proposed yet. Moreover, while taxonomies of interaction environment exist there are currently no formalisation of this concept. Bearing in mind the importance of contextual information that can be obtained from interaction environment and benefits that it can be for development of AUIs, in the current paper we propose a conceptual framework for modelling the environment.

\section{A Contextual Framework for AUIs}
\label{framework}

We propose that the environment of the user can be understood as a System-of-Systems (SoS) which contains sub-systems such as the user, the device which implements the adaptation or personalisation, other users and entities in the same physical space with the user that may have a direct or indirect influence on their behaviour or on the personalisation. Each of these sub-systems posses their own systemic properties; i.e. \textit{components, objectives, relations, behaviour, structure, interface, environment, and functions}~\cite{Maier1996}. Taking such a systemic view allows to better understand the coexisting entities, their inter-dependencies, as well as  their influence on the environment and vice versa. Thus, we can formalise the problem of UI adaptation $p$ as a function of the composing systems, given by
\begin{equation}\label{eqn:PuI}
p = f(u, d, e)
\end{equation}
\noindent where $u$ is the user, $d$ is the device to be adapted through its interface, and $e$ is the environment potentially influencing the user as an independent system as well as containing subsystems itself, such as other people and devices that may have an impact on the user. 

Adapting UIs often entails changing some functionalities such as information content, presentation layout, or distinctiveness of an interface to increase its personal relevance to the user. Doing so requires taking into account not only the user but also environmental constraints and potential influences imposed on the user. This essentially means devising an efficient strategy to understand and reason about dynamic interaction responses of the user and, consequently, adapting the UI to the environmental changes. Implementing this, however, is not a trivial task as it requires to derive efficient representations of the user, the coexisting entities, and the environment's state from high-dimensional sensory inputs, and use these information sources to generalise past experiences to new situations in order to better adapt UIs. 

Such types of challenging tasks are remarkably solved by humans and other animals through a harmonious combination of reinforcement learning (RL) and hierarchical sensory processing systems~\cite{serre2005object, fukushima1982neocognitron}. This in particular has inspired the development of several RL algorithms over the years, cf. Nguyen et al.~\cite{nguyen2020deep}. Early RL algorithms were limited to domains in which useful features could be handcrafted, or to domains with fully-observed low-dimensional state spaces. 

Recently, Deep Q-networks (DQNs) can learn successful policies directly from high-dimensional sensory inputs using end-to-end reinforcement learning~\cite{mnih2017methods}. DQNs have been tested in various complicated tasks and were able to outperform all previous RL algorithms~\cite{silver2016mastering, silver2017mastering}. DQNs have also enabled breakthroughs such as ``AlphaGO''~\cite{chen2016evolution} and``AlphaStar''~\cite{10.1145/3319619.3321894}, which have inspired recent work on AUIs in the context of linear menus~\cite{Todi21_menus}.
These advancements demonstrate the potential of RL to build intelligent agents by giving them the freedom to learn by exploring their environment and make decisions to take actions that maximise a long term reward.

We believe that RL can be highly beneficial to AUIs, as it allows learning through exploration, unlike classic approaches and supervised methods, that  often require large amounts of labelled data and are harder to train with continuous action spaces. Taking this inspiration, in the following we reformulate the goal of AUIs as an RL task by extending the high-level formalisation in \autoref{eqn:PuI}.

In classic RL, agents interact with their environment through a sequence of observations, actions, and rewards~\cite{watkins1992q}. At a given time, an agent takes an observation (i.e., information about the state of the environment) and takes an action that will maximise a long-term reward. The agent then observes the consequence of the action on the state of the environment and the associated reward. It then continues to make decisions about which actions to take in a way that maximises the cumulative future reward. This is done by learning action value function:
\begin{equation}
Q^{*} (s,a) = \max_\pi \mathbb{E}\left [\sum_{t\geqslant 0}^{} \gamma^{t} r_{t}|s_{0}= s, a_{0} = a, \pi  \right ]
\end{equation}

\begin{figure}[!h]
  \centering
  \includegraphics[width=\linewidth]{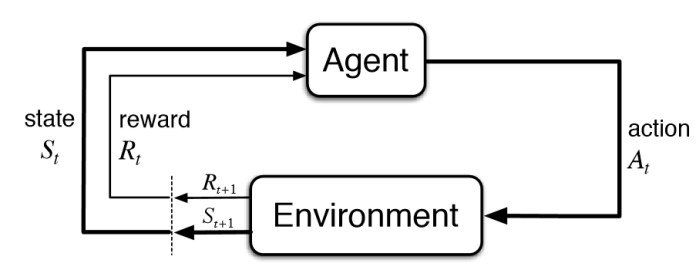}
  \caption{Classic Reinforcement Learning framework~\cite{sutton1998introduction}.} \label{fig:RL}
\end{figure}
which is the maximum sum of rewards $r_{t}$ discounted by $\gamma$ at each time step $t$, achievable by a policy $\pi = p(a|s)$, after making an observation of $s$ and taking an action $a$. This means that RL agents operate based on a policy $\pi$ to approximate $Q$-values (state-action pairs) that maximise a future reward.  \autoref{fig:RL} illustrates the schematics of the different components in classic RL. We refer the reader to the work of Watkins and Dayan~\cite{watkins1992q, sutton1998introduction} for the details on Q-learning and RL.

\begin{figure*}[h!]
\centering
  \includegraphics[width=0.65\linewidth]{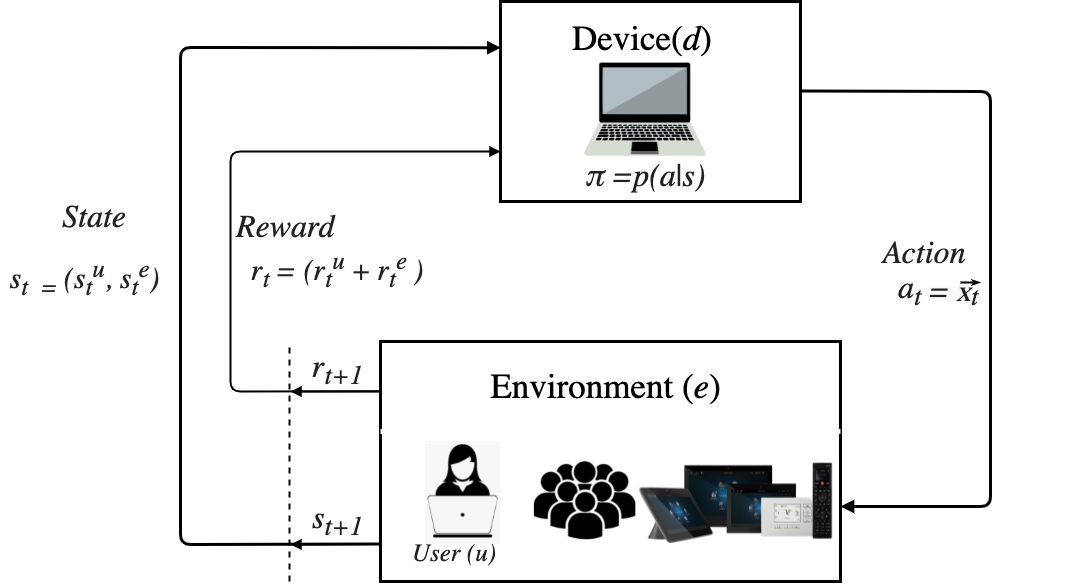} 
  \caption{User interface adaptation as an RL task.}
  \Description{}
  \label{fig:RL_UI}
\end{figure*}

Adopting this to the context of AUIs, the agent corresponds to the device $d$ which operates based on some policy $\pi$. The action taken by the device $a_t$ at any time step $t$ corresponds to selecting an optimal UI configuration layout denoted by \textbf{x} = ($x_{1}$, \dots, $x_{N}$) representing a configuration over $N$ features. The observation state $s_t$ then corresponds to the combination of the state of the user $s_{t}^{u}$ and the state of the environment $s_{t}^{e}$. The reward the device receives $r_t$ for selecting a configuration \textbf{x} is the sum of the rewards deemed appropriate for the corresponding states of the user and the environment: $r_t$ =  $r_{t}^{u}$ +  $r_{t}^{e}$. \autoref{fig:RL_UI} illustrates the problem of AUIs as an RL task.

Following the classic RL formulation, at each step the approximation of the optimal Q-value function $Q^{*}$ will be refined by enforcing the Bellman equation~\cite{watkins1992q}, which can be reformulated for our AUI setting by substituting an action $a$ with the task of selecting a configuration \textbf{x} given by
\begin{equation}
Q^{*} (s,\textbf{x}) =  \mathbb{E}_{s'\sim\varepsilon } \left [ r + \gamma\max_{\textbf{x}'} Q^{*} (s',\textbf{x}')|s,\textbf{x} \right ],
\end{equation}
which states that given any state-configuration pair $s$ and \textbf{x}, the maximum cumulative reward achieved is the sum of the reward for that pair, $r$, plus the value of the next state we end up with, $s'$. The value at state $s'$ will be the maximum over all possible configurations $\textbf{x}'$ at $Q^{*} (s',\textbf{x}')$. Thus, the optimal policy $\pi^{*}$ corresponds to selecting the best UI configuration in any state, as specified by $Q^{*}$. 

In this iterative process, the Bellman equation is used as a value-iteration algorithm to refine $Q^{*}$:
\begin{equation}
Q_{i+1} (s,\textbf{x}) =  \mathbb{E}\left [ r + \gamma\max_{x'} Q_{i} (s',\textbf{x}')|s,\textbf{x} \right ],
\end{equation}
where $Q_{i}$ converges to $Q^{*}$ as $i$ approaches to infinity.

In the context of AUIs, we are interested in finding an optimal policy on which the device operates in order  to select the best possible configuration given the states of the user and the environment. Since users in such settings experience cognitive and physical workload, they respond differently depending on: individual skills, UI familiarity, preferences, etc. In particular, these user states may correspond to implicit interactions which are often hard to detect and analyse. Nevertheless, considerable advances have been made in inferring emotional, cognitive, and behavioural states through response monitoring of various biosignals~\cite{dinh2020stretchable}, for example electrocardiography~\cite{Cairns16_ecg}, electroencephalography~\cite{Holler17_eeg}, or eye-gaze~\cite{Menges19_eye} and mouse~\cite{Bruckner21_choice} movements. Thus, such biosignals can be leveraged as an underlining technique of our approach to iteratively infer user states to find the best personalised UI configuration through an optimal policy given the inferred user states.
\vspace{-2mm}

\section{Use Case: Learning Management Systems}
\label{case study}

With the COVID-19 pandemic, the use of learning management systems has increased dramatically~\cite{szopinski2022student}. Since these systems can be accessed with different devices (e.g. desktop, mobile, tablet) and in various environments (e.g. home-office, conference rooms, common areas in universities), it is important that the UI can consider these environments for adaptation. For the purpose of demonstrating our proposed framework, we will consider a use case where an AUI accounts for changes in environment in order to improve the user's remote learning experience.

Mark is a student participating in a remote lecture that is broadcasted live on an online learning platform. 
Mark is attending from his apartment (environment(\textit{e})), which he shares with other students. 
When the lecture starts, Mark joins from his living room using a laptop 
and listens to the audio via laptop speakers, as he does not have any headset available. The current layout of the content presentation on his laptop (device (\textit{d})) corresponds to a configuration (\textbf{x}) chosen by the AUI.  
Some time later, a persistent drilling noise starts coming out from his neighbour's flat,
so the AUI automatically activates closed captions. 
The captions appear to be too small to read, which makes Mark lean closer to the screen. 
The AUI detects this movement and modifies the font size to improve readability. 
After 20 minutes, the drilling noise stops and the QA session starts, 
so Mark switches on his microphone to ask a question. 
As he begins to speak, suddenly one of his flatmates enters the living room. 
The AUI recognises a visual change in the environment and automatically blurs Mark's background 
to minimise the distracting impact that it may have on other students who are currently attending the lecture. 
Shortly after the doorbell rings. 
Mark is now alone in the apartment so he should answer the door. 
However, since the QA session is still going on, Mark continues listening to it 
via his mobile phone as he walks towards the door. 
The AUI modifies the layout to account for the limitations of the small screen display. 
The lecture concludes as Mark opens the door: it is a deliveryman with the headphones that he recently ordered. 

In the presented scenario the changes in the environment such as the drilling noise (acoustic signal), someone entering the room (background visual signal), walking towards the door (posture change), etc. are signals collected from the user as well as the environment that serve as input to interpret the state of user ($s_{t}^{u}$) and environment ($s_{t}^{e}$). In RL, rewards guide the exploratory nature of an agent. Hence, in our proposed framework, the device expects some form of feedback from the user and the environment in the form of a reward (\textit{r}), either positive or negative, for choosing the content presentation (\textbf{x}) given their observed states. It should be noted that modelling such a reward is rather a challenging task due to difficulties in interpreting the true meaning of multi-modal signals collected from the users and their environment.
Nonetheless, the above use case illustrates how modelling and responding to the changes in  user environment can contribute to an improved learning experience by using signals capture from the environment to automatically adapt user interface to different interaction circumstances.

\section{Current Challenges and Implications}
\label{implications}

Although our framework paves a path towards incorporating the environment in the design of AUIs, 
the RL formulation suffers from a scalability problem. 
This is due to the fact that one must compute $Q(s,\textbf{x})$ for every state-configuration pair 
in order to select the best UI configuration. This is computationally infeasible as the configuration space is potentially infinite, only limited by the number of elements that can be adapted. Take for example the domain of web applications: with CSS and JavaScript it is possible to modify any UI element at will~\cite{Leiva11_ace}.

Recent research in RL has addressed the scalability problem by using function approximators, 
typically a neural network, to estimate the action-value function $Q(s,a;\theta)$ $\approx$ $Q^{*}(s,a)$ 
where $\theta$ is the trainable parameters (weights) of the neural network~\cite{mnih2017methods}. 
Deep Q-learning is one of the most commonly used techniques to approximate optimal action-value functions using a neural network. Hence, we can define our Q-function approximator using a neural network too. This means that, in the forward pass while training the network, we use a loss function to minimise the error of the Bellman equation, thereby determining how far $Q(s,\textbf{x})$ is from the target $Q^{*}(s,\textbf{x})$, given by
\begin{equation}\label{eqn:Qv}
L_{i}(\theta_{i})  =  \mathbb{E}_{s,\textbf{x}\sim \rho(.) } \left [y_{i} - Q(s,\textbf{x};\theta_{i}) \right ]^{2}
\end{equation}
where,
$y_{i} =  \mathbb{E}_{s'\sim\varepsilon } \left [ r + \gamma\max_{\textbf{x}'} Q(s',\textbf{x}';\theta_{i-1})|s,\textbf{x} \right ]$.
Then, then backward pass is a gradient update with respect to the Q-function parameters $\theta$. 

It is also evident from recent works that there are different variations of DQN that have enjoyed a huge success in RL tasks such as Actor-Critic methods~\cite{xiang2019continuous}, which combine DQN with Deep Deterministic Policy-Gradient Algorithms (DDPG)~\cite{lillicrap2015continuous} and a multi-agent version of actor-critic methods~\cite{lowe2017multi, ryu2020multi}. We are optimistic that our proposed framework, complemented by such techniques, may overcome scalability issues and open new, interesting research opportunities in AUIs, as it enables adaptation by learning through exploration. 

Overall, our proposed contextual framework can bring several practical benefits. 
Firstly, it allows a contextual modelling approach of AUIs that incorporates environment and accounts for its constituent elements. Secondly, framing the problem as an RL task opens possibility of multidisciplinary research at the intersection of Machine Learning, Cognitive Science and Psychology. Thirdly, if applied in practise, the RL architecture empowers us to learn complex interaction responses of users and their environment through exploration, which, in turn, can lead to design of more seamless and user-friendly AUIs. Furthermore, although we have used a single user setting to illustrate the proposed architecture, it could be extended to more complicated interaction scenarios such as multi-user setting and group adaptation, where single adaptation is applied to multiple users.

Nevertheless, we would like to note that there are several open questions that need to be addressed before the practical implementation of our framework becomes feasible. Firstly, determining the reward function poses a challenge as it requires keeping the user in the loop in order to provide their feedback based on system's actions. Secondly, the system needs to correctly interpret and act on a multitude of signals from the environment that  may potentially have contradictory meaning. Thirdly, due to complexity of the environment, the neural network requires to be exposed to a large number of examples which makes training extremely time consuming and negatively impacts scalability of the system. 

We anticipate that by concerted collaborative efforts, the AUI community will begin to address these challenges to pave way to practical implementation of the proposed contextual framework in the future.  

\begin{acks}
This work was supported by the Horizon 2020 FET program of the European Union through the ERA-NET Cofund funding grant CHIST-ERA-20-BCI-001.
\end{acks}

\bibliographystyle{ACM-Reference-Format}
\balance
\bibliography{bibliography.bib}


\end{document}